\documentclass[a4paper,11pt]{article}
\pdfoutput=1
\usepackage{jheppub} 
\usepackage{graphicx,color,rotating}
\usepackage{hyperref}
\usepackage{epsfig,color}
\usepackage{slashed}
\usepackage{amsfonts}
\usepackage{ulem}
\usepackage{color}

\graphicspath{{D:/Desktop/draft}}
\usepackage{tabularx}
\usepackage{graphicx}
\usepackage{adjustbox}
\usepackage{tabularx}

\begin{document}
\title{511 keV line and primordial black holes from first-order phase transitions}

\renewcommand{\thefootnote}{\arabic{footnote}}

\author{
Po-Yan Tseng$^{1}$, and Yu-Min Yeh$^{1}$}
\affiliation{
$^1$ Department of Physics, National Tsing Hua University,
101 Kuang-Fu Rd., Hsinchu 300044, Taiwan R.O.C. \\
}

\date{\today}

\abstract{511 keV gamma-ray excess from the Galactic center is a long lasting anomaly
without satisfying astrophysical explanation.
  Hawking evaporation of hypothetical primordial black hole (PBH) with mass $1.0\times 10^{-17} \lesssim M_{\rm PBH}/M_\odot \lesssim 8.0\times 10^{-17}$
  and fractional abundance $10^{-3} \lesssim f_{\rm PBH} \lesssim 1.0$,
  gives rise substantial non-relativistic electrons/positrons
  annihilating into diphoton, well reproduces the 511 keV line.
  However, it is obscure of the mechanism behind to form PBH with
  meteoritical mass in the early Universe.
  In this work, we investigated the production mechanism of PBHs through
  a cosmological first-order phase transition induced by quartic effective
  thermal potential via a scalar field in dark sector.
  We found the phase transition with vacuum energy,
$\mathcal{O}(1)\lesssim B^{1/4}/{\rm MeV} \lesssim \mathcal{O}(100)$,
  produces the desired monochromatic PBH mass and abundance fraction.
  Correlated signatures of gravitational wave and extragalactic gamma-ray,
  respectively coming from phase transition and black hole evaporation,
  are within $\mu$Ares and AMEGO/e-ASTROGAM/COSI/XGIS-THESUS projected sensitivities.
  Finally, we include the PBH mass function from FOPT
  and found it can not improve the explanation to 511 keV excess.
}

\maketitle

\section{Introduction}
    Primordial black hole (PBH), alternative to particle oriented dark matter (DM), represents one of the macroscopic DM candidate forming in the early Universe~\cite{Hawking:1971ei,Chapline:1975ojl,Khlopov:2008qy,Carr:2016drx,Carr:2020gox,Carr:2020xqk,Green:2020jor}.
Suppose PBH is around tenth of solar mass, it provides convincing explanation~\cite{Clesse:2016vqa,Bird:2016dcv,Sasaki:2016jop} for LIGO/VIRGO~\cite{LIGOScientific:2016aoc,LIGOScientific:2016sjg,LIGOScientific:2017bnn} gravitational wave (GW) signals from black hole mergers.
However, the PBH mass is determined by the production mechanism, and several models have been proposed.
After inflation for example, PBHs can be generated through the collapse
of the overdensity regions developed from the primordial fluctuation re-entering the horizon, which makes the PBH mass comparable to the horizon mass~\cite{Carr:1974nx,Sasaki:2018dmp}.
Other papers discussed PBHs directly form following the bubble collisions of first-order phase transition (FOPT)~\cite{Hawking:1982ga,Moss:1994iq,Konoplich:1999qq,Kodama:1982sf,Gross:2021qgx,Baker:2021nyl}.
Nevertheless, these mechanisms are inefficient to generate sub-solar mass PBH.

In this work, we consider an alternative PBHs production mechanism from FOPT in the early Universe.
During FOPT,
an intermediate state dubbed as Fermi balls (FBs) were composed by
the dark fermions $\chi$'s which were filtered by the expanding bubbles and trapped in the false vacuum~\cite{Hong:2020est,Marfatia:2021twj,Witten:1984rs,Bai:2018dxf}.
The filtering can be realized by introducing a dark scalar $\phi$ couples to dark fermion, where the dark scalar develops a vacuum expectation value in the true vacuum and generates mass differences of dark fermion between the true and false vacua, so that it forces the dark fermions stay in false vacuum if they do not have sufficient thermal energy.
Furthermore, the attractive Yukawa interaction associating with dark scalar triggers a FB collapsing into PBH~\cite{Kawana:2021tde}.
Since the Standard Model (SM)
electroweak and QCD (quantum chromodynamics) phase transitions are smooth crossovers,
we therefore invoke the FOPT being generated in the dark sector, which can also be established via the same dark scalar.
Specifically, a commonly discussed
quartic effective thermal potential of a scalar field $\phi$ is adopted.
In addition, the scalar field coupling to the $\chi$ through Yukawa interaction serves two purposes.
First, the non-zero VEV (vacuum expectation value) of the scalar increases the mass $\chi$ in the true vacuum, due to the energy-momentum conservation at bubble wall, which keeps the $\chi$'s in the false vacuum
once the mass difference is larger than the critical temperature of the FOPT.
For the other purpose, the Yukawa interaction induces attractive force between $\chi$'s
and triggers the collapsion of a FB into PBH.
This instability of FB happened when the range of Yukawa interaction becomes comparable with the mean separation distance of $\chi$'s
in a FB~\cite{Kawana:2021tde}.
The produced PBH mass is associated with the volume of false vacuum bubble at percolation temperature of FOPT and in general can be much smaller than the horizon mass.
Therefore, meteoritic PBH mass can be easily produced from vast parameter space.
In additional to the monochromatic PBH mass, this scenario also predicts PBH mass function originating from the radii distribution of the false vacuum bubbles~\cite{Lu:2022paj}.
We will consider both the monochromatic and mass function distributions of PBH, and discuss their implications for the results.

From the astrophysical observation perspective,
an excess amount of photons with energy 511 keV from the central region of Milky Way has been confirmed by SPI/INTEGRAL
~\cite{Bouchet:2010dj}.
If someone identified this as a result of electron-positron annihilation,
it requires the injection rate of $2\times 10^{43}$ non-relativistic positrons per second
~\citep{Weidenspointner:2004my,Churazov:2004as,Weidenspointner:2007rs,Jean:2005af,Prantzos:2005pz}.
Various potential astrophysical sources have been proposed for the excess,
but encountered difficulties explaining the characteristic of the signal~\cite{Prantzos:2010wi}. In addition, the MeV scale DM annihilation is hardly consistent with cosmic microwave background observation~\cite{Wilkinson:2016gsy}.
Recently, it has been suggested that the positrons responsible for the 511 keV line excess
might be produced through the Hawking evaporation
~\cite{Hawking:1975vcx,Gibbons:1977mu} of PBHs
~\cite{Keith:2021guq,Frampton:2005fk,Bambi:2008kx,Cai:2020fnq,Laha:2019ssq,DeRocco:2019fjq}.
Assuming the PBHs spacial distribution follows the NFW halo profile
and concentrates in the inner Galaxy.
The PBH mass of $M_{\rm PBH}\sim \mathcal{O}(5\times 10^{16})~{\rm g}$ and the $\mathcal{O}(10^{-3})$ fraction of total DM abundance, has been shown,
not only produces the right amount of positron flux
but also be consistent with existing limits of
COMPTEL/INTEGRAL gamma-ray observations~\cite{Keith:2021guq}.
In addition, the near future MeV gamma-ray observations, e.g. AMEGO~\cite{Fleischhack:2021mhc,Ray:2021mxu}, e-ASTROGAM~\cite{e-ASTROGAM:2017pxr}, COSI~\cite{Caputo:2022dkz}, and XGIS-THESUS~\cite{Ghosh:2021gfa},
will provide better sensitivities to explore these PBHs.
Our analysis mainly focus on AMEGO/e-ASTROGAM.

In this work, we aim to find the corresponding parameters space under the FOPT scenario, which generates the PBH mass and relic abundance to explain the 511 keV excess.
As a result, the correlated gamma-ray spectra form PBH evaporation and GW signals from FOPT are predicted.
This paper is organized as following: In section~\ref{sec:PBH_511keV}, base on NFW and isothermal distributions,
we scrutinize the preferred PBH mass and abundance fraction for 511 keV excess.
The realization of PBH formation and correlated signals production
are discussed in section~\ref{sec:PBH_FOPT} and \ref{sec:BP}, respectively.
Section~\ref{sec:mass_function} includes the mass function and discusses its influences on the results.
Finally, we summarize the results in section~\ref{sec:conclusion}.

\bigskip

\section{PBH evaporation}
\label{sec:PBH_511keV}
    Hawking emission describes a PBH thermally produce primary particles with masses lighter than the PBH temperature $T_{\rm PBH}=M^2_{\rm Pl}/M_{\rm PBH}$, which is numerically formulated as
    \begin{eqnarray}
    T_{\rm PBH}\simeq 5.3~{\rm MeV}\times \left( \frac{10^{-18}M_\odot}{M_{\rm PBH}} \right)\,.
    \end{eqnarray}
    The secondary particle productions come from the decays or fragmentation
    of primary particles (e.g $e^\pm,\mu^\pm,\pi^\pm,\pi^0$).
    A PBH dominantly emits photons and neutrinos,
    but for $M_{\rm PBH}/M_\odot \lesssim 10^{-16}$, the Hawking temperature is high enough,
    PBH can copiously produces electrons and positrons.
    The emission rate of primary particle $i$ is given by~\cite{Hawking:1974rv,Hawking:1975vcx}
    \begin{eqnarray}
    \frac{dN_i}{dE dt}=
    \frac{n^{\rm d.o.f}_i  \Gamma_i(E,M_{\rm PBH})}{2\pi (e^{E/T_{\rm PBH}}\pm 1)}\,,
    \end{eqnarray}
    where $n^{\rm d.o.f}_i$ indicates the degrees of freedom of $i$ particle,
    and the fermions (bosons) are distinguished by the $+(-)$ in the denominator.
    The graybody factor $\Gamma_i(E,M)$ varies from particle to particle
    and is derived from considering a wave packet scattering
    in the PBH spacetime geometry from the PBH horizon to an observer
    at infinity.

    Numerically, the particle spectra of PBH evaporation can be take into account by
    the software package BlackHawk v2.1~\cite{Arbey:2019mbc,Arbey:2021mbl}
    to compute the $\gamma$ and $e^\pm$ production rates,
    including both primary and secondary components of PBH evaporation.
    The positron emission rate per single PBH can be numerically obtained via integrating out the evaporation spectrum
    \begin{eqnarray}
    L_{e^+}= \int dE\, \frac{dN_{e^+}}{dE dt}\,.
    \end{eqnarray}
    Notice that BlackHawk encounters the PBH mass evolution due to Hawking evaporation, so that, in general, the particle spectra are time-dependent.

\bigskip

\subsection{Galaxy center 511 keV line}

\begin{figure}[t]
      \centering
      \includegraphics[width=13cm]{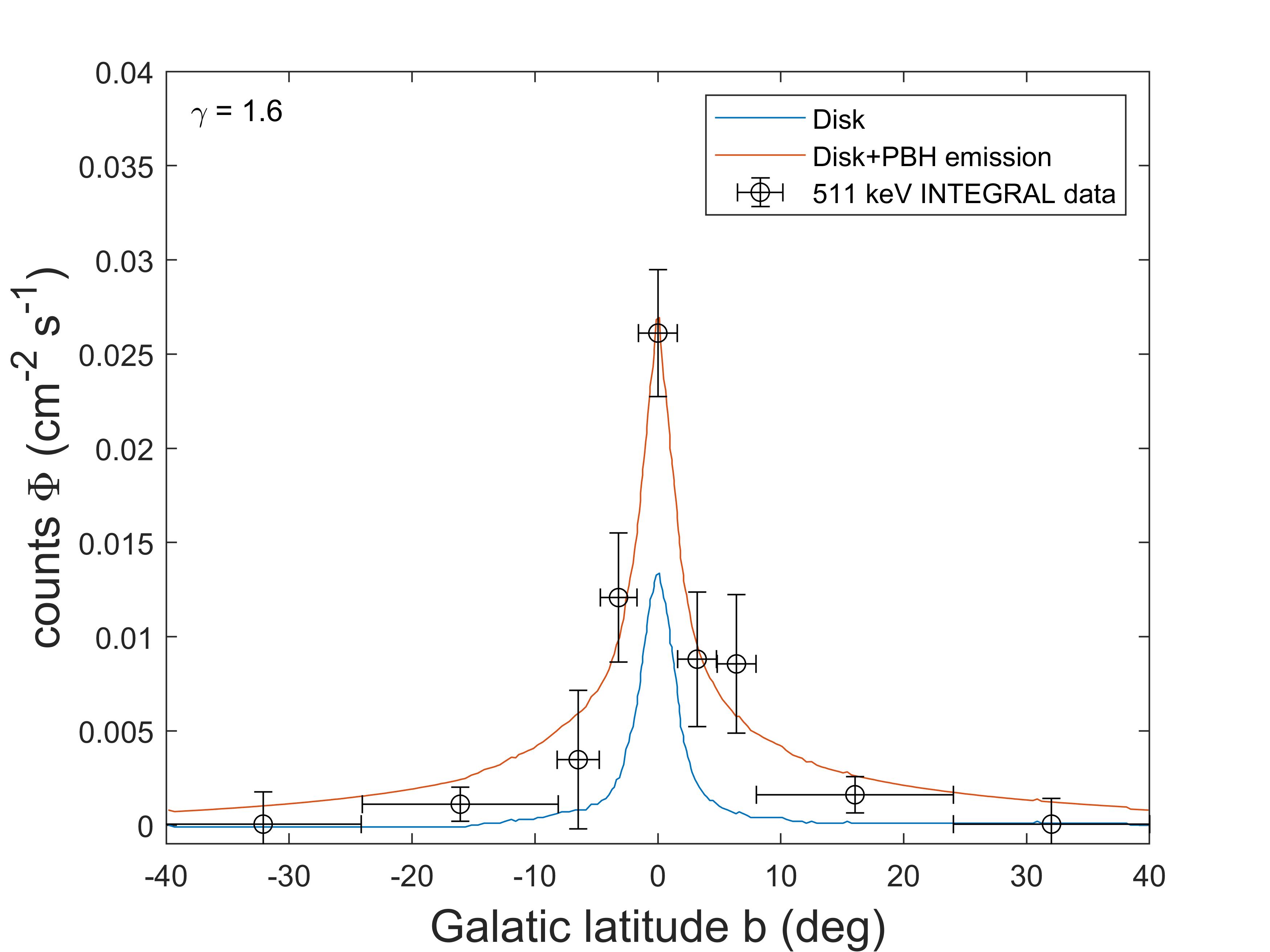}
      \caption{The orange curve is combined 511 keV gamma-ray flux
      from PBH evaporation with $M_{\rm PBH}=10^{16}$ g, $f_{\textrm{PBH}}=10^{-4}$ and astrophysical source~\cite{Robin:2003uus}(blue curve).
      Comparing to the INTEGRAL data from 508.25 keV to 513.75 keV~\cite{Bouchet:2010dj}}\label{fig:1}
    \end{figure}

    The galaxy center 511 keV line observed by SPI/INTEGRAL~\cite{Bouchet:2010dj}
     exceeds the astrophysical contributions,
     and the originating mechanism behind remains unclear.
    In this paper, we focus on the explanation of PBHs in mass range from $10^{15}\,$g
    to $2\times 10^{17}\,$g, which substantially produce particles through Hawking evaporation.
Positrons are abundantly emitted from a PBH and more than 95\% of them become non-relativistic via ionization. Among the non-relativistic positrons, fraction of $(1-f)$ directly annihilate with electrons to produce two 511 keV photons, while the others fraction of $f\approx0.967$ form a positronium bound state ~\cite{Keith:2021guq}. Among the later case, 25\% of positroniums annihilate to form pair of 511 keV photons, and the rest 75\% yield three photons with energy less than 511 keV. Turn out, the average number of 511 keV photons produced per positron is
    \begin{eqnarray}
    2(1-f)+\frac{2f}{4}\approx0.55\,,
    \end{eqnarray}
    and thus contributes to the 511 keV photons flux from the near galactic center region,
    covering solid angle $\Delta\Omega$, is formulated as
    \begin{eqnarray}
    \label{eq:511_photon}
    \Phi_{\mathrm{PBH}}(\Delta\Omega)=\frac{0.55L_{e^+}(M_{\textrm{PBH}})f_{\textrm{PBH}}}{4\pi M_{\textrm{PBH}}}\int_{\Delta\Omega}{\int_{\textrm{l.o.s}}{\rho(\ell,\Omega)d\ell}d\Omega} \,,
    \end{eqnarray}
    where $L_{e^+}(M_{\textrm{PBH}})$ is the positron production rate from a PBH, $f_{\textrm{PBH}}$ be the fraction of PBHs abundance to the DM relic abundance. The integrals are performed over solid angle and along the line-of-sight.
    We assume the PBHs spacial distribution follows the Navarro-Frenk-White (NFW)~\cite{Navarro:1995iw} and Isothermal~\cite{Bergstrom:1997fj} halo profiles parametrized as~\cite{Bertone:2004pz}
    \begin{eqnarray}
    \rho(r)=\frac{\rho_0}{(r/R_s)^\gamma[1+(r/R_s)^\alpha]^{(\beta-\gamma)/\alpha}}\,,
    \end{eqnarray}
    where $r$ is the distance from the galactic center,
    and the normalization parameter $\rho_0$ adjusts the DM density near our solar system,
    $\rho=0.4$ $\textrm{GeV}/\textrm{cm}^3$ at $r=8.25$ kpc.
    For NFW, we fix $(\alpha,\beta,\gamma)=(1.0,3.0,1.6)$ with scale radius $R_s=20$ kpc in this analysis, meanwhile we choose $(\alpha,\beta,\gamma)=(2.0,2.0,0)$ and $R_s=3.5$ kpc for Isothermal profile.
    Combining with the galactic disk contribution,
    the 511 keV line predicted from PBHs emission, depending on the angular distribution of galactic latitude $b$ and averaging over the longitude profile $-8^\circ<\ell<+8^\circ$, is shown in Fig.~\ref{fig:1},
    which is compared with the INTEGRAL data of photon energy from 508.25 keV to 513.75 keV~\cite{Bouchet:2010dj}.

\begin{figure}[t]
      \centering
      \includegraphics[width=13cm]{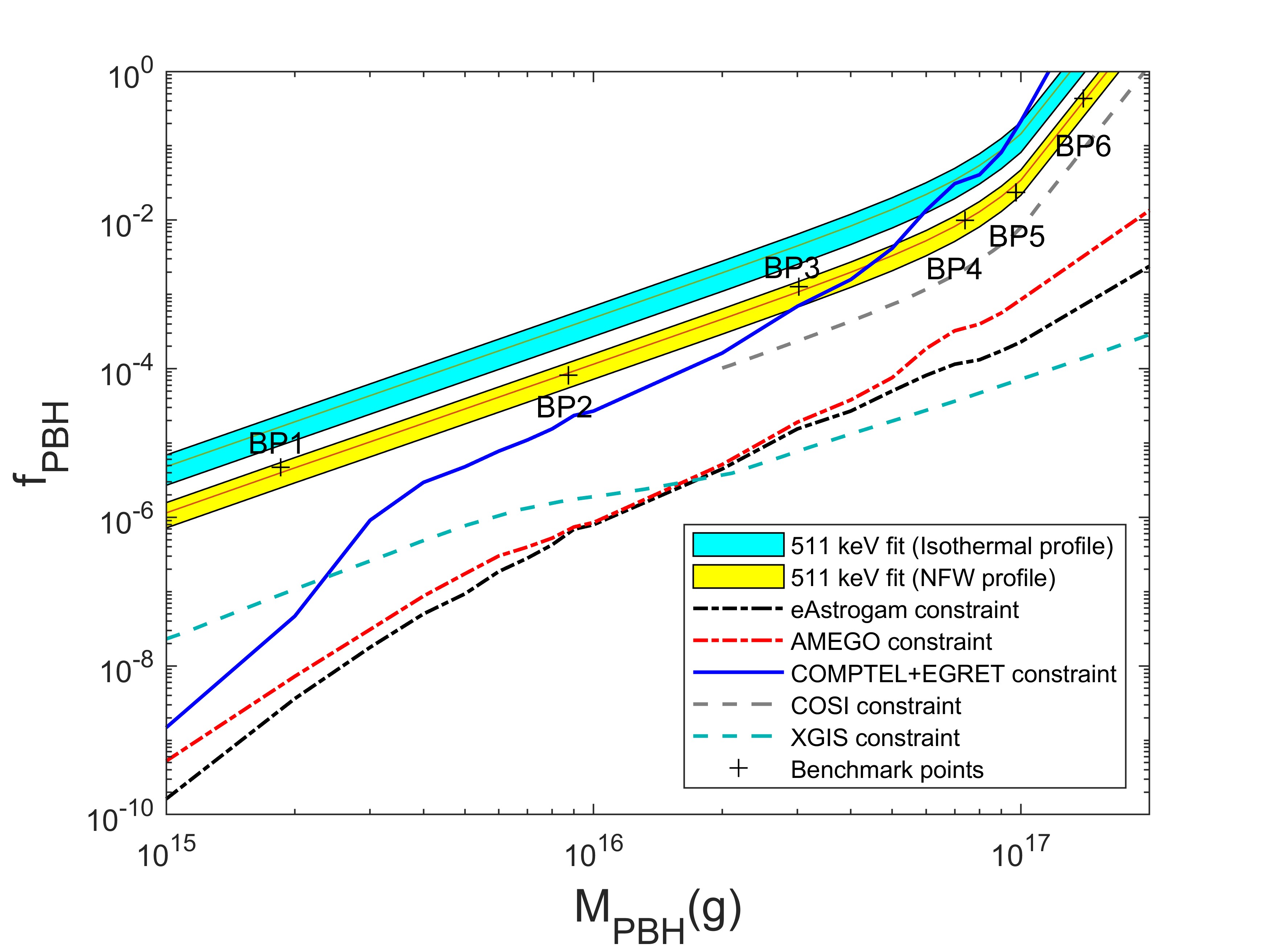}
      \caption{The yellow and cyan bands,
      which assume the NFW and Isothermal profiles respectively, can produce the 511 keV excess observed by INTEGRAL via the PBH evaporation.
      The constraints of COMPTEL+EGRET and sensitivities of eASTROGAM and AMEGO are shown by the solid curves, requiring the extragalactic photon from PBHs contribution does not exceed the observation data.
      The benchmark points ({\bf BP}s) label the PBHs produced from the FOPT scenario, and their parameters are listed in Table~\ref{table:1}.
      The prospect sensitivity curves of COSI~\cite{Caputo:2022dkz} and XGIS-THESUS~\cite{Ghosh:2021gfa} are also shown for comparison.
      }\label{fig:2}
    \end{figure}

    The best fit of PBH fraction to the 511 keV INTEGRAL data is determined by the $\chi^2$ statistic test, which is defined by
    \begin{eqnarray}
    \chi^2(f_\mathrm{PBH})=\sum_{i}{\left[\frac{\Phi_{s}(b_i,f_\mathrm{PBH})+\Phi_{\mathrm{disk}}(b_i)-\Phi_{\textrm{data}}(b_i)}{\sigma_{\Phi}(b_i)}\right]^2}\,,
    \end{eqnarray}
    and $i$ runs over the INTEGRAL data points in Fig.\ref{fig:1}.
    $\Phi_{\mathrm{disk}}$ stems from astrophysical positron emission in the disk. $\Phi_{s}$ represents the PBHs contribution after taking into account the angular resolution of INTEGRAL, which is obtained by Gaussian smearing as
    \begin{eqnarray}
    \Phi_s(b_i,f_\mathrm{PBH})=\int^{}_{}{ \frac{1}{\sqrt{2\pi \sigma^2_{b_i}}}\exp{\left[-\frac{(b_i-b)^2}{\sigma^2_{b_i}}\right]}\,\Phi_{\mathrm{PBH}}(b,f_\mathrm{PBH})db} \,,
    \end{eqnarray}
    where $b_i$ and $\sigma_{b_i}$ respectively correspond to the central value and the horizontal error of each INTEGRAL data point.
    The yellow region of Fig.~\ref{fig:2} shows
    the best-fit parameter region of $(M_{\rm PBH},f_{\rm PBH})$,
    which requires $\chi^2-\chi_{\mathrm{min}}^2\leq4$.
    For $M_{\rm PBH}\simeq 1.6\times 10^{17}~{\rm g}$,
    PBHs can serve as 100\% DM abundance
    and explain the 511 keV line excess simultaneously.
    The $M_{\rm PBH}\lesssim 3\times 10^{16}~{\rm g}$
     overproduces the extragalactic gamma-ray which will be discussed in following sub-section
    and thus is excluded by current upper bound of EGRET in accordance with Fig.~\ref{fig:2}.


    \bigskip

\subsection{Extragalactic and Galactic gamma-rays}

    The PBH evaporation also produces primary photons and contribute to extragalactic gamma-ray background, hence the current gamma-ray observational data can be converted into the upper limit of the PBHs abundance.
    In order to find the constraint in the parameter space, we calculate the PBH extragalactic photon contributions and compare with the current experimental data of COMPTEL/EGRET/FermiLAT~\cite{Fermi-LAT:2018pfs}, as well as future sensitivities of AMEGO/e-ASTROGAM~\cite{e-ASTROGAM:2017pxr,Fleischhack:2021mhc,Laha:2020ivk}.
    The extragalactic photon flux, integrating the PBH evaporation from different red-shift, is given by~\cite{Marfatia:2021hcp}
    \begin{eqnarray}
    \label{eq:EG_photon}
    \frac{d\Phi^{\rm EG}}{dE}=\frac{1}{4 \pi}\, \int^{\mathrm{min}(t_{\mathrm{eva}},t_0)}_{t_\mathrm{CMB}}{c[1+z(t)]\frac{f_\mathrm{PBH}\rho_{\mathrm{DM}}}{M_\mathrm{PBH}}\frac{d^2N_\gamma}{d\tilde{E}dt}\bigg|_{\tilde{E}=[1+z(t)]E}dt}\,,
    \end{eqnarray}
where $\rho_\mathrm{DM}=1.27\mkern5mu \mathrm{GeV }\mathrm{m}^{-3}$ is the average dark matter density in the extragalactic medium. From this expression,
we see the extragalactic photon flux is independent from the uncertainties of DM density profile near the galactic center,
thereby it provides more robust prediction than that from the Galactic center gamma-ray.
The lower limit of temporal integral starts at the time of last scattering, $t_\mathrm{CMB}=3.8\times10^5\mkern5mu\mathrm{yr}$, after that, photons decouple from thermal plasma and become free steaming. Photons generated before recombination epoch abruptly thermalize with other particles and never escape to Earth. Meanwhile, the upper limit of integral is set by the shorter period of the lifetime of PBH ($t_\mathrm{eva}$) or the age of Universe ($t_0=13.77\times10^9\mkern5mu\mathrm{yr}$). Applying matter-dominated approximation to the evolution of Universe, the time-redshift relation can be approximated as
$1+z(t)=\left(t_0/t\right)^{\frac{2}{3}}$.
For the galactic component, the photon flux is expressed by~\cite{Ray:2021mxu,Calabrese:2021src}
\begin{eqnarray}
    \label{eq:GA_photon}
    \frac{d\Phi^{\rm GA}}{dE}=\frac{1}{\Delta \Omega}\, \frac{f_\mathrm{PBH}}{4 \pi M_\mathrm{PBH}}
    \frac{d^2N_\gamma}{dEdt}\int_{\Delta \Omega} \int_{\rm l.o.s} \rho(\ell,\Omega)\, d\ell \, d\Omega\,,
\end{eqnarray}
where one of the major uncertainty comes from the DM density distribution near the Galaxy Center.
    For different masses of PBH,
    we show the upper limits of $f_\mathrm{PBH}$ in Fig.~\ref{fig:2},
    which are compatible with the present COMPTEL/EGRET/ upper bounds (blue curve)
    and future AMGEGO/e-ASTROGAM gamma-ray sensitivities (red/black curves).

\bigskip

\section{First-order phase transition}
\label{sec:PBH_FOPT}

According to above analysis, the preferred PBH mass is around $M_{\rm PBH}\simeq 10^{17}~{\rm g}$, which is smaller than solar mass by several orders of magnitude, and thus must not be produced from standard astro process,
therefore we are going to discuss a novel production mechanism through FOPT in early Universe.
The PBH can be produced through a two-step processes during FOPT
which is induced by the effective thermal potential of a scalar $\phi$ in the hidden sector.
The FB firstly forms from the aggregation of dark fermions $\chi$'s and plays the role as intermediate state before PBH formation.
Secondly, because the $\chi$'s inside a FB couple through $\phi$, which provides an attractive Yukawa potential.
The length of Yukawa interaction increases as FB temperature decreases,
and thus Yukawa energy eventually dominates the total FB energy
which causes the FB to collapse into a PBH~\cite{Kawana:2021tde}.


    Specifically, the formation of PBHs can be realized by the Lagrangian~\cite{Marfatia:2021hcp}
    \begin{eqnarray}
    \mathcal{L}\supset  \bar{\chi}(i\slashed{\partial}-M_i)\chi  -g_{\chi}\phi \bar{\chi}\chi-V_{\mathrm{eff}}(\phi,T)\,,
    \end{eqnarray}
    where $M_i$ represents the bare mass of $\chi$ in the false vacuum,
    $V_{\mathrm{eff}}$ is the finite-temperature quartic effective potential
    commonly appears from theoretical models
    ~\cite{Dine:1992wr,Adams:1993zs}
    \begin{eqnarray}
    V_{\mathrm{eff}}(\phi,T)=D(T^2-T^2_0)\phi^2-(AT+C)\phi^3+\frac{\lambda}{4}\phi^4\,,
    \end{eqnarray}
    which induces the cosmological FOPT.
    The FOPT happened when the temperature becomes lower than the critical temperature $T_c$,
    which is determined by the condition $V_{\rm eff}(0,T_c)=V_{\rm eff}(v_\phi(T_c),T_c)$,
    at this moment, the false vacuum ($\langle \phi \rangle=0$)
    was tunnelling to the true vacuum ($\langle \phi \rangle = v_\phi$).
    We define the parameter $B$ to be the zero-temperature potential energy density difference
    between the false and true vacuum~\cite{Marfatia:2021twj}.
    Finally, the quartic potential can be described by these input parameters~\cite{Marfatia:2021twj}
    \[ \lambda,  A,  B,  C,  D\,. \]

    We use the analytical expression for the Euclidean action $S_3(T)/T$
    of quartic potentials~\cite{Adams:1993zs}
    to compute the bubble nucleation rate per unit volume $\Gamma(T)$
    and then obtain the fraction of space in the false vacuum $F(t)$~\cite{Marfatia:2021twj}.
    The phase transition temperature $T_\star$ identified as percolation
    temperature can be obtained when a fraction $1/e$ of the space
    remains in the false vacuum. Such that the corresponding time of phase transition $t_\star$
    is given by $F(t_\star)=1/e\simeq 0.37$.

In general, a cosmological FOPT is accompanied with a GW signal from bubble collisions, sound waves, and Magnetohydrodynamic turbulence.~\cite{Huber:2008hg,Nakai:2020oit,Espinosa:2010hh,Caprini:2015zlo,Kehayias:2009tn}.
In our case, since the bubble wall reaches a relativistic terminal velocity and corresponds to a non-runaway bubble, the sound waves dominate the GW production~\cite{Caprini:2015zlo}. We follow the semi-analytical approaches to evaluate the GW spectra~\cite{Marfatia:2020bcs,Caprini:2015zlo}.
    The GW characteristic frequency is related to
    the
    inverse time duration of the FOPT
    $\beta/H_\star\simeq T_\star\,
    \left[d(S_3/T)/dT\right]$~\cite{Nakai:2020oit},
    which is normalized to Hubble time scale. In addition to this parameter,
    the GW spectrum also depends on $\alpha$, the strength of FOPT,
    which is defined as the ratio between
    false vacuum energy to the total radiation energy.
A large uncertainty of predicted GW signal may come from the bubble wall velocity $v_w$.
We focused on the non-ultrarelativistic limit and adopted the Chapman-Jouguet velocity~\cite{Steinhardt:1981ct} to predict the GW spectrum in most of our analysis.
Since the Chapman-Jouguet condition is generally not fulfilled ~\cite{Espinosa:2010hh,Laine:1993ey},
we parameterize the uncertainty in the GW signal via varying $v_w$ around the Chapman-Jouguet velocity.
In Fig.~\ref{fig:4}, the shaded band shows the uncertainties of GW spectra by varying the bubble wall velocity in the range, $0.3 \leq v_w \leq 1.0$.
We can observe that the lower bubble wall velocity suppresses the GW amplitude meanwhile increases the peak freqency.

    In order to find $n_{\rm FB}$ the number density of FB, we need to know
    the critical volume $V_\star$ which is defined from a volume in false vacuum satisfying the
    condition $\Gamma(T_\star)V_\star R_\star\simeq v_w$~\cite{Hong:2020est},
    and thus this volume would not further divide into smaller one during its shrinking.
    The $R_\star$
    is the corresponding radius of the critical volume.
    Under the assumption that each critical volume corresponds to one FB,
    the volume fraction of false vacuum implies the relation
    $n_{\rm FB}|_{T_\star}= F(t_\star)/V_\star$ at $t_\star$.

    Inside a FB, to avoid complete annihilation $\bar{\chi} \chi \to \phi \phi$,
    there must be a nonzero number density asymmetry
    $\eta_{\rm DM} \equiv (n_\chi-n_{\bar{\chi}})/s$ (normalized to the entropy density).
    Then we define $Q_{\rm FB}$ the total number of $\chi$ forming a FB:
    $Q_{\rm FB}\equiv \eta_{\rm DM}(s/n_{\rm FB})|_{T_\star}$~\cite{Hong:2020est}.

    In general, the dark and SM sectors need not to be in thermal equilibrium,
    otherwise the dark radiation ($\phi$ and $\chi$) would contribute to
    the effective number of extra neutrino species
    $\Delta N_{\rm eff}$
    and exceed the current observational upper bound.
    In fact, lowering the dark sector temperature than SM sector
    alleviates the tension between $\Delta N_{\rm eff}$ and observations~\cite{Marfatia:2021twj}.
    Thus, we include the temperature ratio $r_T\equiv T_\star/T_{\rm SM_{\star}}$
    at $t_\star$ as one of the input parameter for FOPT.

    The dark fermion particles $\chi$'s aggregated to form
    macroscopic FBs due to the FOPT.
    Subsequently, the attractive Yukawa force between $\chi$'s mediated by
    $\phi$ destabilizes the FBs to form PBHs.
    In the following section, we quantitatively discuss these processes.


  \bigskip

 \subsection{FB properties and PBH formations}
    The detail derivations for the FB mass and radius, we follow Ref.~\cite{Marfatia:2021hcp}. The total energy of a FB, including the Fermi gas kinetic energy; Yukawa potential energy; and the temperature-dependent potential energy difference between the false and true vacua, is approximately written as

    \begin{eqnarray}
        \label{eq:FB_energy}
        E_{\rm FB}&=&
        \frac{3\pi}{4} \left( \frac{3}{2\pi} \right)^{2/3} \frac{Q^{4/3}_{\rm FB}}{R}
        \left[1+\frac{4 \pi}{9}\left(\frac{2\pi}{3} \right)^{1/3} \frac{R^2 T^2}{Q^{2/3}_{\rm FB}}
        \left(1+\frac{3}{2\pi^2}\frac{M^2_i}{T^2} \right)
         \right] \nonumber \\
        &&
        -\frac{3 g^2_\chi}{8 \pi} \frac{Q^2_{\rm FB} L^2_\phi}{R^3}
        + \frac{4 \pi}{3} V_0 R^3 \left(1+\frac{T^2 M^2_i}{12 V_0} \right) \,, 
    \end{eqnarray}
    where $V_0(T)\equiv V_{\rm eff}(0,T) - V_{\rm eff}(v_\phi(T),T)$, which at zero temperature returns into $B$, and $M_i$ associates with the bare mass of $\chi$. Because the FB is a macroscopic object, we ignore the contribution from the surface tension.
    Inside a FB, $\chi$'s self-coupling is through the attractive Yukawa term $g_{\chi}\phi \bar{\chi}\chi$, and the length of Yukawa interaction is given as
    \begin{eqnarray}
     L_\phi(T)\equiv \left(\frac{d^2 V_\mathrm{eff}}{d\phi^2}\bigg|_{\phi=0}\right)^{-1/2}=(2D(T^2-T^2_0))^{-1/2} \,.
    \end{eqnarray}

    Finding $R_{\rm FB}$ for the stable radius of FB, we require $dE_{\rm FB}/dR=0$,
    which yields a cubic polynomial equation with three solutions.
    The largest value among the three solutions
    gives the FB radius $R_{\rm FB}$.
    Substituting $R_{\rm FB}$ back to $E_{\rm FB}$ gives rise the mass of the FB, then
    we have
    \begin{eqnarray}
        R_{\rm FB} &=& \left[ \frac{1}{4}\left( \frac{3}{2\pi} \right)^{2/3} \frac{Q^{4/3}_{\rm FB}}{V_0} \right]^{1/4} X^{1/2} \\
        M_{\rm FB} &=& \frac{3}{4} Q_{\rm FB} \left( 9 \pi^2 V_0 \right)^{1/4} X^{-3/2} \nonumber \\
        &\times& \left\lbrace X+ \frac{4}{9}\left(1+\frac{T^2 M^2_i}{12V_0} \right)X^3
        + \left(  \frac{2\pi T^2}{9 V^{1/2}_0}+ \frac{M^2_i}{3\pi V^{1/2}_0} \right) X^2
        - \frac{2g^2_\chi L^2_\phi V^{1/2}_0}{3\pi}\right\rbrace\,,
        \label{eq:FB_mass_radius}
    \end{eqnarray}
    where denote
    \begin{eqnarray}
        X&\equiv& \left(1+\frac{T^2 M^2_i}{12V_0} \right)^{-1}  \nonumber\\
        & \times &\Bigg[\left( 1+ \frac{13}{108}\frac{T^2M^2_i}{V_0}+ \frac{\pi^2}{81} \frac{T^4}{V_0} + \frac{1}{36\pi^2}\frac{M^4_i}{V_0} \right)^{1/2} \cos\theta-\frac{\pi}{18} \frac{T^2}{V^{1/2}_0}\left(1+\frac{3}{2\pi^2}\frac{M^2_i}{T^2} \right) \Bigg]\,. \nonumber
       \end{eqnarray}
To make a FB stable in the false vacuum, it further requires the conditions~\cite{Kawana:2021tde}
\begin{eqnarray}
\label{eq:FB_stable}
\frac{dM_{\rm FB}}{dQ_{\rm FB}} \leq g_\chi v_\phi+M_i\,,~~~~\frac{d^2M_{\rm FB}}{dQ^2_{\rm FB}}\leq 0\,,
\end{eqnarray}
where the first inequality holds dark fermions inside FB, whereas the second condition is automatically satisfied by the surface tension to prevent FB fission into smaller pieces.
In our region of interest, non-zero $M_i$ is always necessary to fulfil the first condition with perturbative Yukawa coupling
$|g_\chi|\leq \sqrt{4 \pi}$.

    If we neglect the Yukawa energy, $X$ reduces into
    \begin{eqnarray}
        X
        &\simeq& \frac{\sqrt{3}}{2}\left(1-\frac{1}{4\sqrt{3}\pi}\frac{M^2_i}{V^{1/2}_0}- \frac{\pi}{6\sqrt{3}}\frac{T^2}{V^{1/2}_0}-\frac{5}{216} \frac{M^2_iT^2}{V_0}  \right)\,.\,
    \end{eqnarray}
    and the assumption $V_0\gg T$ and $M_i$, Eq.~(\ref{eq:FB_mass_radius}) can be simplified into
    \begin{eqnarray}
        \label{eq:FB_mass_radius_simple}
        R_{\rm FB}&=& \left[ \frac{3}{16} \left( \frac{3}{2\pi} \right)^{2/3}
        \frac{Q^{4/3}_{\rm FB}}{V_0} \right]^{1/4}
        \left(1-\frac{1}{8\sqrt{3}\pi}\frac{M^2_i}{V^{1/2}_0}
        - \frac{\pi}{12\sqrt{3}}\frac{T^2}{V^{1/2}_0}-\frac{5}{432} \frac{M^2_iT^2}{V_0} \right)\,,  \\
        M_{\rm FB} &=& Q_{\rm FB}\left( 12 \pi^2 V_0 \right)^{1/4}
        \left(1+ \frac{\sqrt{3}}{8\pi} \frac{M^2_i}{V^{1/2}_0}
        +\frac{\pi}{4\sqrt{3}}\frac{T^2}{V^{1/2}_0}-\frac{1}{16} \frac{M^2_i T^2}{V_0} \right)\,.
    \end{eqnarray}

Due to the fact that the magnitude of Yukawa energy
    \begin{eqnarray}
        |E_Y|\simeq \frac{3g^2_\chi}{8 \pi} \frac{Q^2_{\rm FB}}{R}\,  \left(\frac{L_\phi}{R} \right)^2
    \end{eqnarray}
increases as the temperature decreases, at temperature $T_\phi$ defined as Yukawa energy and Fermi energy equality, the FB becomes unstable and starts collapsing to a PBH. This roughly coincides when $L_\phi$ equals to the mean separation distance of $\chi$'s, i.e., $L_\phi\simeq R_{\rm FB}/Q^{1/3}_{\rm FB}$.
Therefore, there are two scenarios of PBH formation:
i) If $T_\phi \leq T_\star$, the FB forms as an intermediate state before collapsing into PBH.
ii) If $T_\phi > T_\star$, the $\chi$'s enclosed in a critical volume directly collapse into PBH without FB formation.
In the first scenario, we can adopt the above formulas of FB to compute the mass of PBH, i.e. $M_{\rm PBH}(T_\phi)=M_{\rm FB}(T_\phi)$, and the
number density follows the adiabatic evolution of Universe $n_{\rm PBH}|_{T_\phi}=n_{\rm FB}|_{T_\star} s(T_\phi)/s(T_\star)$~\cite{Kawana:2021tde}.
Consequently, the PBH relic abundance and fraction at present Universe is given by
\begin{eqnarray}
\Omega_{\rm PBH}h^2 = \frac{M_{\rm PBH}|_0 n_{\rm PBH}|_0}{3 M^2_{\rm Pl}(H_0/h)^2}\,,~ f_{\rm PBH}\equiv \frac{\Omega_{\rm PBH}h^2}{\Omega_{\rm DM}h^2}\,,
\end{eqnarray}
where the Hubble constant $H_0=2.13h\times 10^{-42}~{\rm GeV}$.
In the following section,
we will select the benchmark points belong to the first scenario.

\bigskip

\section{Correlated signals}
\label{sec:BP}

    \begin{figure}[t]
      \centering
      \includegraphics[width=7.5cm]{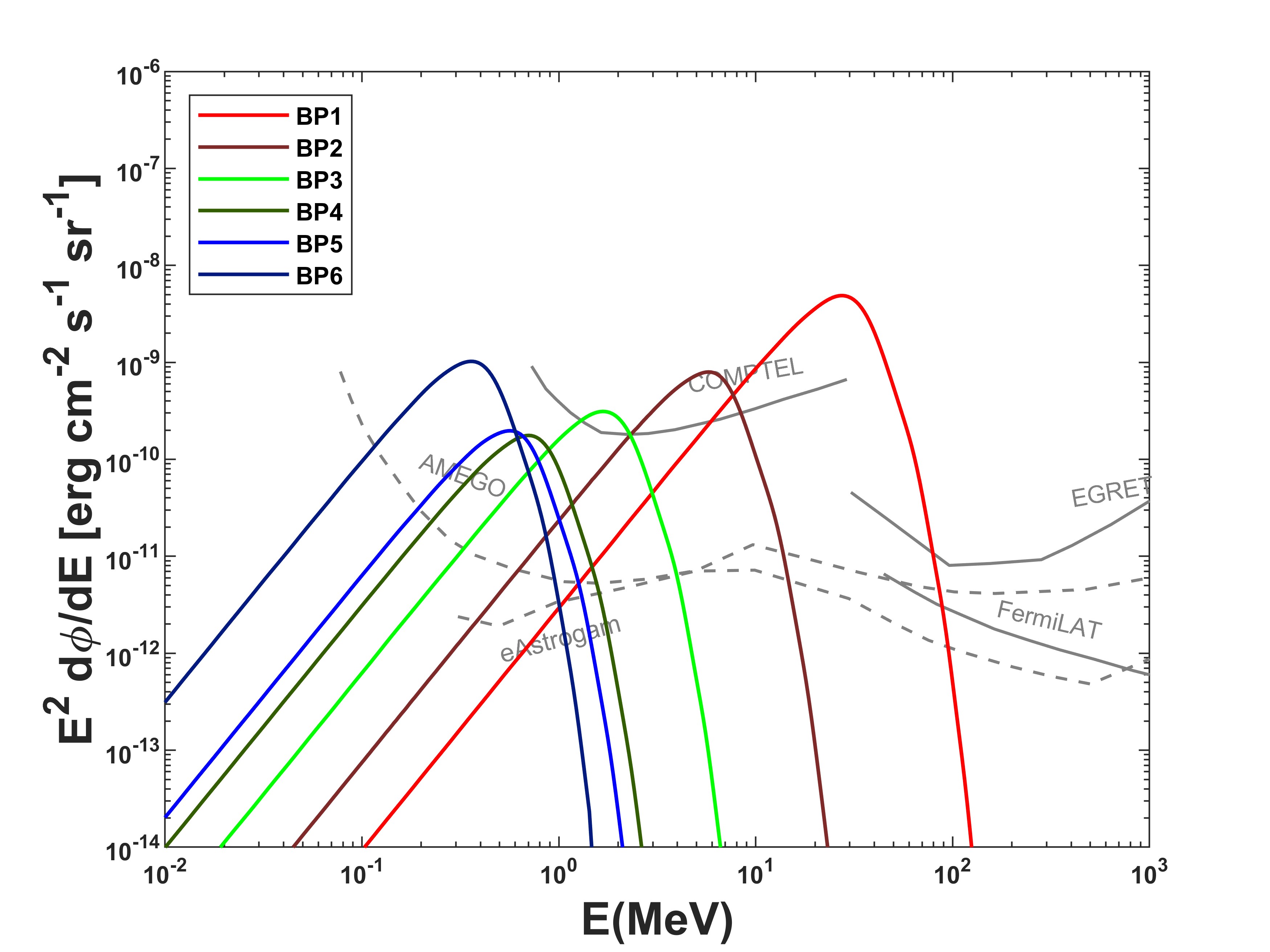}
      \includegraphics[width=7.5cm]{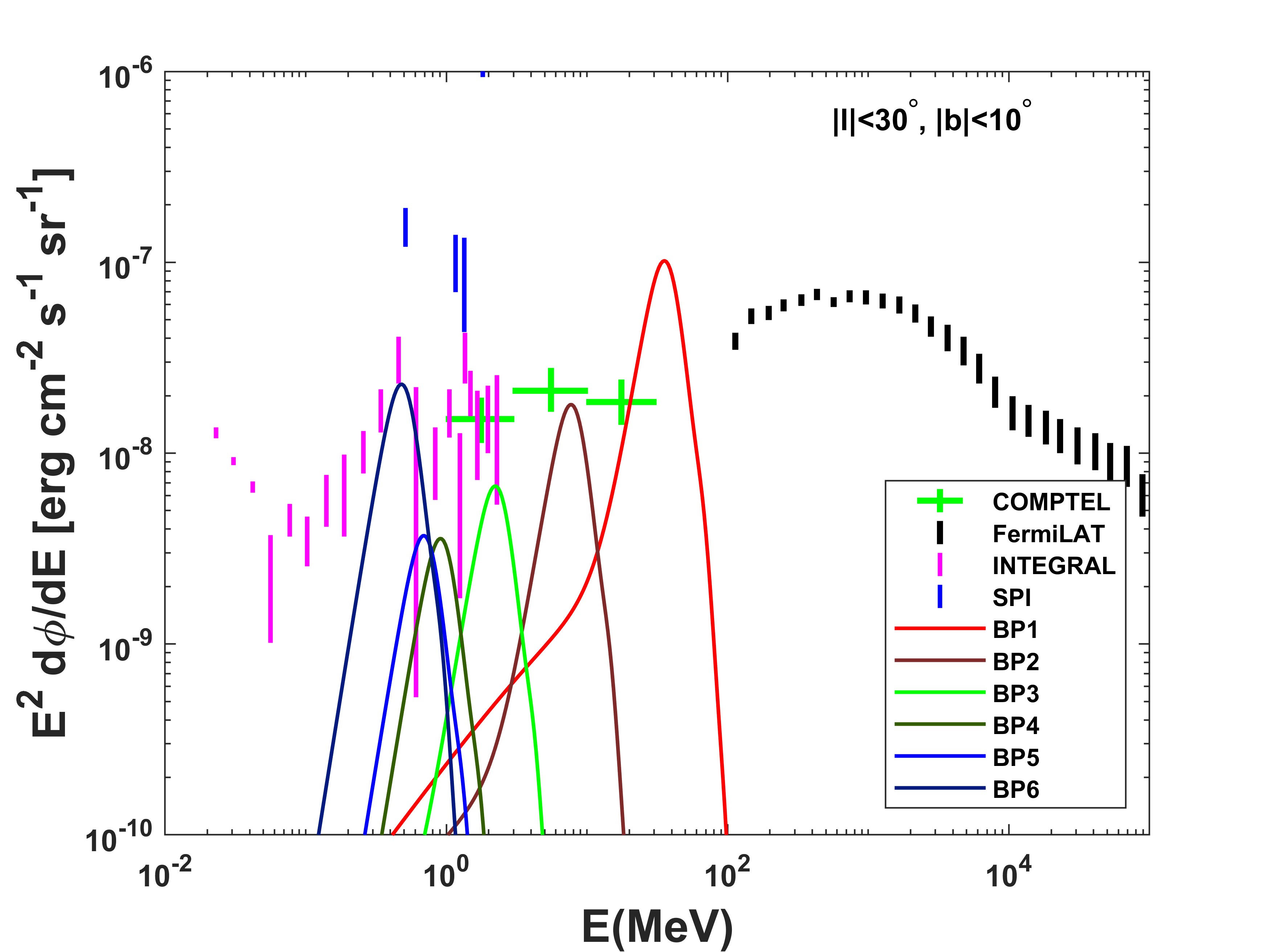}
      \caption{The predicted photon fluxes from the {\bf BP}s listed in Table~\ref{table:1} for extragalactic (left-panel) and inner Galactic (right-panel). The current limits from COMPTEL/EGRET/FermiLAT~\cite{Fermi-LAT:2018pfs} and projecting sensitivities of AMEGO/e-ASTROGAM~\cite{e-ASTROGAM:2017pxr,Fleischhack:2021mhc,Laha:2020ivk} are shown in gray curves in left-panel, meanwhile, in the right-panel, the {\bf BP}s fluxes are compared with the INTEGRAL/SPI/COMPTEL/FermiLAT inner Galactic data ($|l|< 30^\circ$, $|b|<10^\circ$)~\cite{e-ASTROGAM:2017pxr}.}\label{fig:3}
    \end{figure}

     \begin{figure}[t]
      \centering
      \includegraphics[width=13cm]{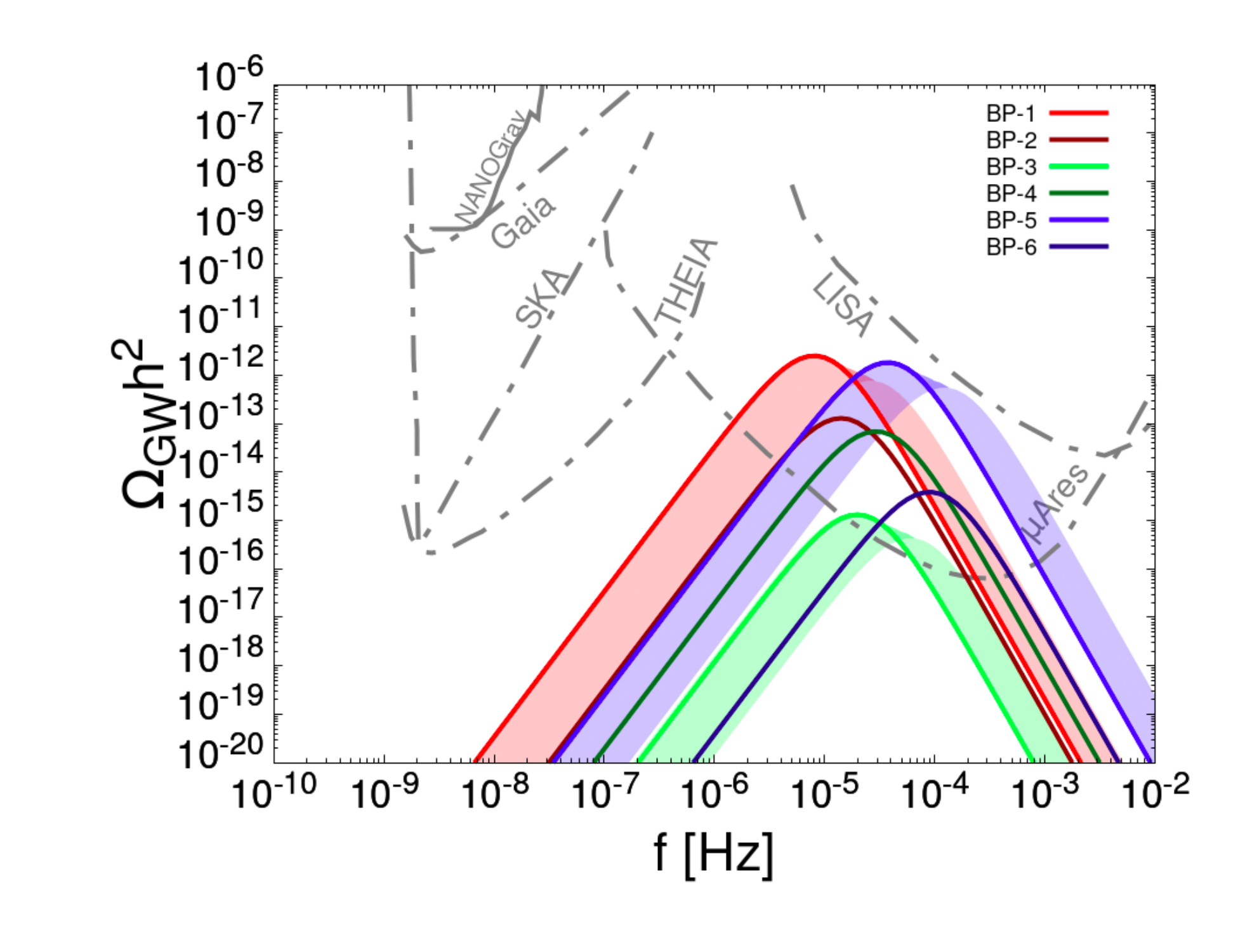}
      \caption{
      Gravitational wave power spectra for {\bf BP}s in Table~\ref{table:1}.
      The shaded bands show the uncertainty in the GW spectra for {\bf BP-1}, {\bf BP-3}, and {\bf BP-5} by varying the bubble wall velocity in the range, $0.3 \leq v_w \leq 1.0$.
      }\label{fig:4}
    \end{figure}

    Since the FOPT behaves as common origin for both PBHs and GW,
    it predicts the correlated signals among the 511 keV line,
    extragalactic photon spectrum,
    and GW spectrum.
    More specifically, the quartic potential dictates the FOPT
    and then the Yukawa interaction determines the timing of PBHs formation.
    Therefore, we select and scan the input parameters: including the coefficients of the effective potential,
    the asymmetry parameter $\eta_\mathrm{DM}$,
    the temperature ratio of the dark and SM sectors $r_T$,
    the Yukawa coupling $g_\chi$, and the bare DM mass $M_i$,
    in the ranges
\begin{eqnarray}
&& 0.1\leq B^{1/4}/{\rm MeV}\leq 10^4, \mkern15mu 0.1\leq D\leq 10, \mkern15mu 0.05\leq \lambda \leq 0.2 \nonumber \\
&& 0.01\leq C/{\rm MeV} \leq 10^4 , \mkern15mu 0.3\leq r_T\leq 1, \mkern15mu 0.01\leq g_\chi \leq \sqrt{4\pi} \nonumber \\
&& 10^{-3} \leq M_i/B^{1/4} \leq 10\,,
\end{eqnarray}
    where $A$ is fixed to be $0.1$.
    We pick six benchmark points (BPs) satisfying the 511 keV excess with the corresponding PBH mass and $f_\mathrm{PBH}$ are listed in Table~\ref{table:1} and Fig.~\ref{fig:2}.
    \begin{table}[h]
        \centering
        \resizebox{\textwidth}{!}{
        \begin{tabular}{c|c c| c c| c c }
            \hline
            \hline
            & \bf{BP-1} & \bf{BP-2} & \bf{BP-3} & \bf{BP-4}& \bf{BP-5} & \bf{BP-6} \\
           \hline
           \hline
            $B^{1/4}/\rm MeV$ & $15.6123$ & $25.3490$ & $4.8842$ & $26.3187$ & $79.4670$ & $47.7322$ \\
            $\lambda$ & $0.1693$ & $0.0888$ & $0.0999$ & $0.0795$ & $0.0685$ & $0.1903$\\
            $D$ & $1.9103$ & $0.5360$ & $3.5249$ & $1.6074$ & $1.3413$ & $0.8706$ \\
            $\eta_{\mathrm{DM}}$ & $4.70\times10^{-14}$ & $3.60\times10^{-13}$ & $6.70\times10^{-11}$ & $7.60\times10^{-11}$ & $4.35\times10^{-11}$ & $1.08\times10^{-9}$ \\
            $r_T$ & $0.4377$ & $0.3512$ & $0.3149$ & $0.3845$ & $0.3891$  & $0.4707$\\
            $C/\rm MeV$ & $1.5257$ & $1.3976$ & $0.0920$ & $0.5830$ & $2.8082$ & $3.6441$\\
            $g_{\chi}$ & $1.2728$ & $1.1556$ & $1.0275$ & $1.1958$ & $0.9141$ & $1.3278$\\
            $M_i/B^{1/4}$ & $0.5570$ & $0.0516$ & $0.4973$ & $0.0047$ & $0.3048$ & $1.4341$ \\
            \hline
            $M_\mathrm{PBH}/M_{\bigodot}$ & $9.32\times 10^{-19}$ & $4.39\times 10^{-18}$ & $1.52\times 10^{-17}$ & $3.72\times 10^{-17}$ & $4.90\times 10^{-17}$ & $7.04\times 10^{-17}$\\
            $f_\mathrm{PBH}$ & $4.74\times 10^{-6}$ & $8.20\times 10^{-5}$ & $0.0013$ & $0.0100$ & $0.0235$ & $0.4350$\\
          \hline
                    $T^\ast/{\rm MeV}$ & $5.7420$ & $17.9707$ & $1.3860$ & $10.4642$ & $31.3314$ & $30.2029$\\
          $\beta/H$ & $4558.39$ & $1992.35$ & $31579.08$ & $7762.03$ & $3398.62$ & $8590.15$\\
          $g^\ast$ & $10.7807$ & $14.2988$ & $10.6613$ & $12.0990$ & $16.2574$ & $15.2550$\\
          $U^{1/4}_0/({\rm MeV})$ & $8.4664$ & $21.8516$ & $1.3947$ & $12.2338$ & $53.3528$ & $22.0874$\\
          $v_w$ & $0.9749$ & $0.9514$ & $0.9143$ & $0.9450$ & $0.9852$ & $0.8255$\\
          \hline
        \end{tabular}
        }
        \caption{
        The benchmark points form PBHs after FOPT with $A=0.1$ fixed.
        }
        \label{table:1}
    \end{table}

    According to the {\bf BP}s in Table~\ref{table:1}, in order to produce the desired PBH mass,
    one requires the energy scale $\mathcal{O}(1)\lesssim B^{1/4}/{\rm MeV} \lesssim \mathcal{O}(100)$ of FOPT. By incorporating $\eta_{\rm DM}$, we can obtain the correct value of $f_{\rm PBH}$ to generate 511 keV line.
    Since FB was formed as an intermediate state and treated as a progenitor of PBH, it has to be stable.
    In order to satisfy the stability conditions in Eq.(\ref{eq:FB_stable}), each {\bf BP} has to have large value of $g_\chi$, closing to the perturbative limit,
    and nonzero $M_i$.
    In addition, the observational upper bound of $\Delta N_{\rm eff}\leq 0.5$
    was imposed to all the {\bf BP}s,
    and thereby restricts the temperature ratio to be $0.3\leq r_T\leq 0.5$.
    Here, we assumed the dark sector and SM sector are thermally decoupled,
    such that it suppresses the dark sector contribution to the light degree of freedom.

    The extragalactic and galactic photon contributions of these {\bf BP}s are shown
    in Fig.~\ref{fig:3}.
    Except {\bf{BP-1}} to {\bf{BP-3}}, {\bf{BP-4}} and {\bf{BP-5}} haven't
    been rule out by the present observations COMPTEL/EGRET/FermiLAT, and {\bf{BP-6}} is marginally consistent with INTEGRAL galactic data.
    It is because their associated masses are heavier
    than {\bf{BP-3}} and thus the gamma-ray spectra peak
    at lower energy window between
    0.1 MeV to 1 MeV where the present observations do not have sufficient sensitivities.
    However, this window will be explored
    by future AMEGO/e-ASTROGAM/COSI/XGIS-THESUS observations.
    On the other hand, the corresponding GW spectra of {\bf BP}s, mainly arising from sound waves in the plasma after bubbles collision during the FOPT~\cite{Caprini:2015zlo},
    are shown in Fig.~\ref{fig:4}.
    Their peak frequencies cover the range from $10^{-5}$ Hz to $10^{-4}$ Hz,
    and amplitudes can be substantially explored by future $\mu$Ares~\cite{Sesana:2019vho} telescope.

\bigskip

\section{PBH mass distribution}
\label{sec:mass_function}

    \begin{figure}[t]
      \centering
      \includegraphics[width=13cm]{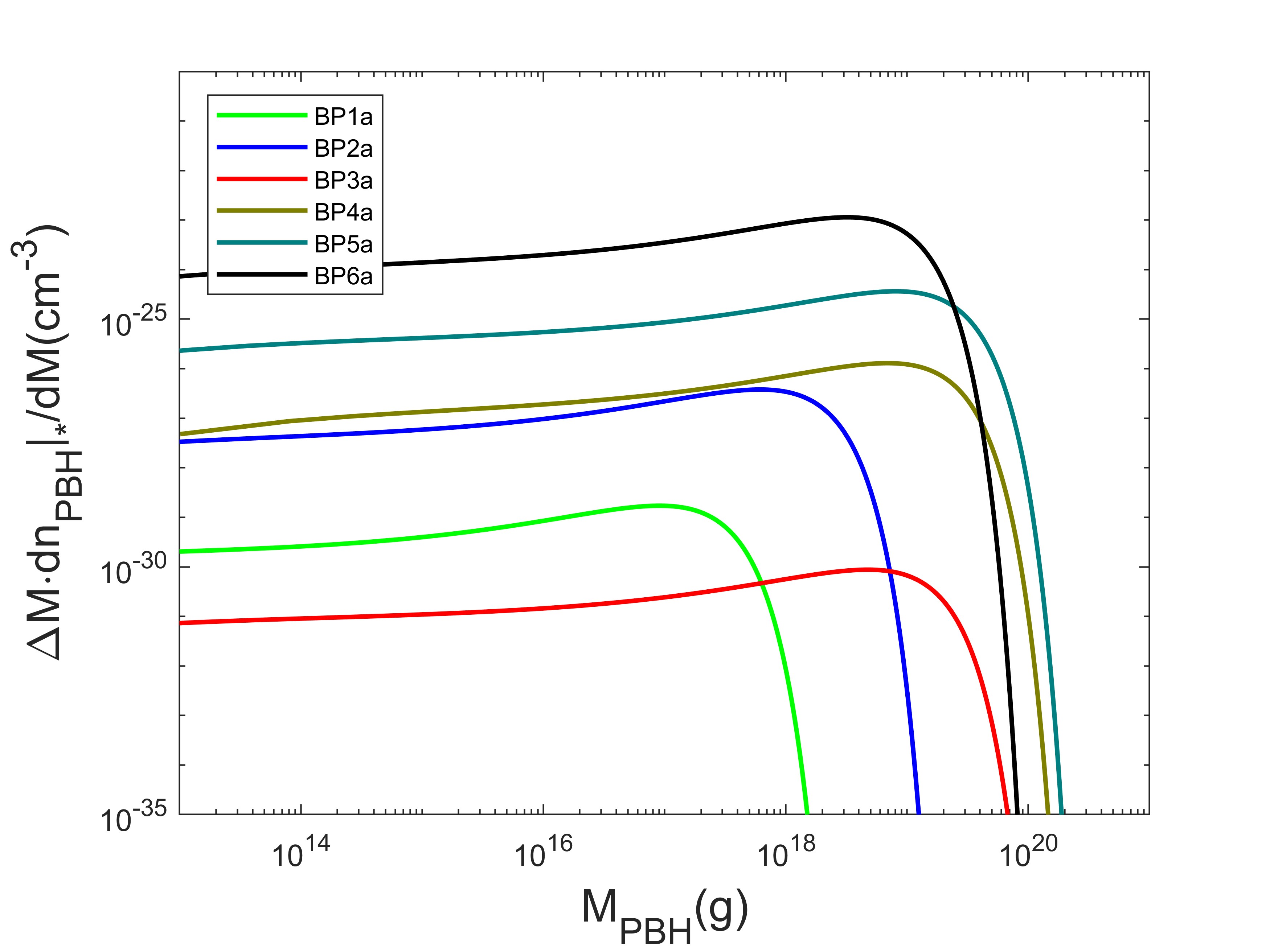}
      \caption{
      The number density distributions of {\bf BP}s from Table~\ref{table:2}.
      }\label{fig:5}
    \end{figure}

\begin{figure}[t]
      \centering
      \includegraphics[width=13cm]{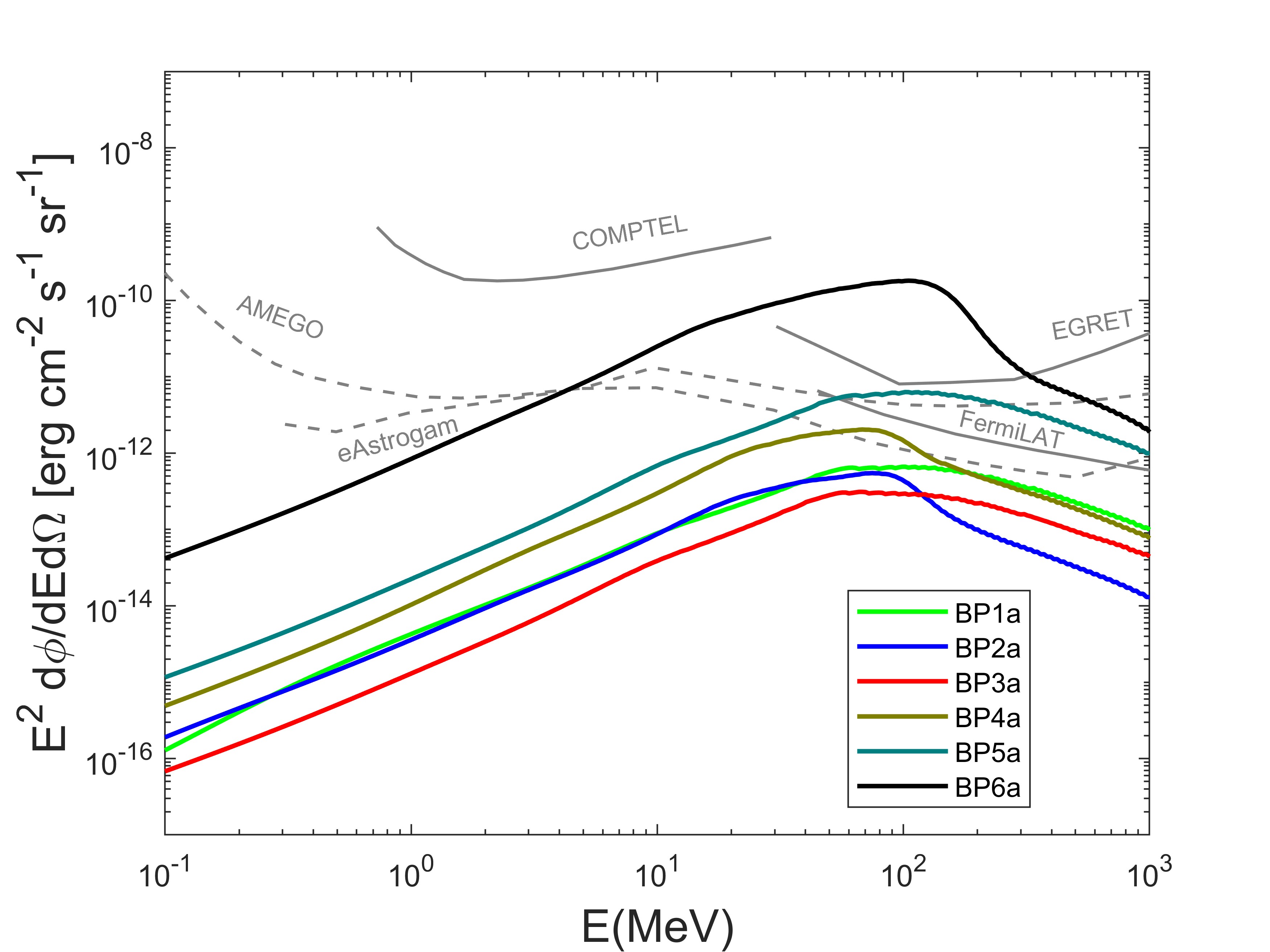}
      \caption{
      The extragalactic photon fluxes of {\bf BP}s with PBH mass distribution from Table~\ref{table:2}.
      }\label{fig:6}
    \end{figure}

\begin{figure}
  \centering
  \includegraphics[width=13cm]{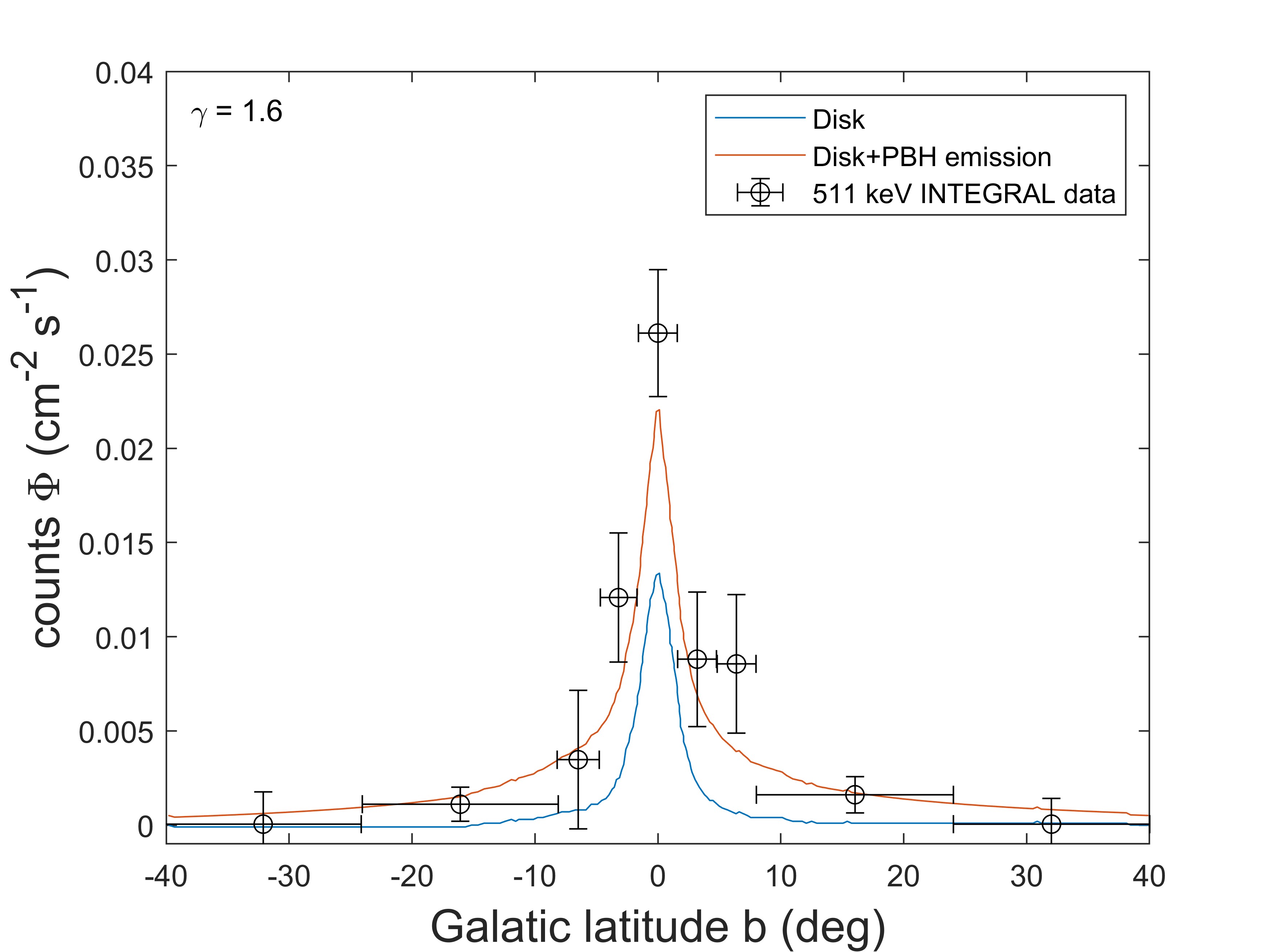}
  \caption{The contribution to 511 keV flux while considering the {\bf BP-6a} PBH mass distribution.}\label{fig:7}
\end{figure}

The above analysis has only considered the monochromatic PBH mass after FOPT. However, because of the radius distribution from the false vacuum bubbles, it induces the mass function of PBH, which may modify the PBH explanation for 511 keV excess.
During the FOPT, the formalism for the number density of false vacuum bubble with respect to its radius distribution at percolation time can be expressed as~\cite{Lu:2022paj}
    \begin{eqnarray}\label{5.1}
\frac{dn_{\rm fv}}{dR_r}(t_\star)\simeq \frac{I^4_\star \beta^4}{192 v^4_w} e^{4\beta R_r/v_w}
e^{-I_\star e^{\beta R_r/v_w}}\times \left(1-e^{-I_\star e^{\beta R_r/v_w}} \right)\,.
\end{eqnarray}
The total number of $\chi$ enclosed in a false vacuum bubble with radius $R_r$ is
\begin{eqnarray}
Q_{\rm FB}=\frac{\eta_\chi s(t_\star)}{f_{\rm fv}(t_\star)}A \frac{4\pi}{3} R^3_r\,,
\end{eqnarray}
where $A$ is a correcting factor to encounter the departure from sphericality of the false vacuum bubble, so that the false vacuum fraction can be normalized as
\begin{eqnarray}
A \int dR_r \frac{4\pi R^3_r}{3} \frac{dn}{dR_r}=0.29
\end{eqnarray}
Since $M_{\rm PBH}\simeq M_{\rm FB}\simeq Q_{\rm FB}(12\pi^2 U_0)^{1/4}$ at leading order expansion~\cite{Marfatia:2021hcp},
we can trade $M_{\rm PBH}$ for $R_r$ via $\frac{dM_{\rm PBH}}{M_{\rm PBH}}=3\frac{dR_r}{R_r}$, and Eq.(\ref{5.1}) can be rewritten as
    \begin{equation}\label{5.4}
        \frac{dn_{\rm PBH}}{dM_{\rm PBH}}= \frac{R_r}{3M_{\rm PBH}} \frac{dn_{\rm fv}}{dR_r}\,,
    \end{equation}
which gives rise the mass function of PBH.
Considering this PBH mass distribution,
the 511 keV and extragalactic photon fluxes from Eq.(\ref{eq:511_photon}) and Eq.(\ref{eq:EG_photon}),
are respectively generalized into
\begin{eqnarray}\label{5.5}
 \Phi^{511\,{\rm keV}}_{\rm PBH} &=& \frac{0.55}{4 \pi} \int dE\,  \frac{n_{\rm PBH}|_0}{\rho_{\rm DM}} \int dM_{\rm PBH}\, \frac{1}{{n_{\rm PBH}}\vert_\star} \frac{d{n_{\rm PBH}}\vert_\star}{dM_{\rm PBH}}
    \frac{dN_{e^+}(M_{\rm PBH})}{dEdt} \int_{\Delta \Omega} \int_{\rm l.o.s} \rho(\ell,\Omega) d\ell d\Omega\,, \nonumber \\ [2mm]
 \frac{d\Phi}{dE}&=& \frac{1}{4 \pi} \int^{\mathrm{min}(t_{\mathrm{eva}},t_0)}_{t_\mathrm{CMB}}dt~
    {c[1+z(t)]\,n_{\rm PBH}|_0
    \int dM_{\rm PBH} \frac{1}{{n_{\rm PBH}}\vert_\star} \frac{d{n_{\rm PBH}}\vert_\star}{dM_{\rm PBH}} \frac{d^2N_\gamma(M_{\rm PBH})}{d\tilde{E}dt}\bigg|_{\tilde{E}=[1+z(t)]E}}\,, \nonumber \\
    \end{eqnarray}
where ${n_{\rm PBH}}\vert_\star$ is the number density at $t_\star$, and $n_{\rm PBH}|_0$ is the number density today.
The relation between ${n_{\rm PBH}}\vert_\star$ and $n_{\rm PBH}|_0$ is given by
    \begin{equation}\label{}
        n_{\rm PBH}|_0=\left(\frac{T_{\rm CMB}}{T_{\star}}\right)^3{n_{\rm PBH}}\vert_\star.
    \end{equation}
We try to find out whether the first-order phase transition parameters shown in Table \ref{table:1} produce the reasonable spectrum of extragalactic gammy-ray and 511keV gammy-ray for the PBH mass distributions.
Even though the averaged mass from mass distribution roughly reproduces the monochromatic result, the mass distribution trends to populate at the lower PBH mass tail, which can be seen from Fig.~\ref{fig:5}. Since the lighter mass PBH evaporates more actively, the {\bf BP}s from Table~\ref{table:1} produce more extragalactic gamma-ray flux than the monochromatic case. In order to suppress their fluxes,
we multiply $B^{1/4}$ and $C$ by $0.04$, which help to lower the $f_{\rm PBH}$ and increase the averaged PBH mass, and other parameters remain the same; so that we select new {\bf BP}s in Table \ref{table:2}.
    \begin{table}[h]
        \centering
        \resizebox{\textwidth}{!}{
        \begin{tabular}{c|c c| c c| c c }
            \hline
            \hline
            & \bf{BP-1a} & \bf{BP-2a} & \bf{BP-3a} & \bf{BP-4a}& \bf{BP-5a} & \bf{BP-6a} \\
           \hline
           \hline
            $B^{1/4}/\rm MeV$ & $0.6245$ & $1.0140$ & $0.1954$ & $1.0527$ & $3.1787$ & $1.9093$ \\
            $\lambda$ & $0.1693$ & $0.0888$ & $0.0999$ & $0.0795$ & $0.0685$ & $0.1903$\\
            $D$ & $1.9103$ & $0.5360$ & $3.5249$ & $1.6074$ & $1.3413$ & $0.8706$ \\
            $\eta_{\mathrm{DM}}$ & $4.70\times10^{-14}$ & $3.60\times10^{-13}$ & $6.70\times10^{-11}$ & $7.60\times10^{-11}$ & $4.35\times10^{-11}$ & $1.08\times10^{-9}$ \\
            $r_T$ & $0.4377$ & $0.3512$ & $0.3149$ & $0.3845$ & $0.3891$  & $0.4707$\\
            $C/\rm MeV$ & $0.0610$ & $0.0559$ & $0.0037$ & $0.0233$ & $0.1123$ & $0.1458$\\
            $g_{\chi}$ & $1.2728$ & $1.1556$ & $1.0275$ & $1.1958$ & $0.9141$ & $1.3278$\\
            $M_i/B^{1/4}$ & $13.9246$ & $1.2899$ & $12.4300$ & $0.1170$ & $7.6200$ & $35.8521$ \\
          \hline
                    $T^\ast/{\rm MeV}$ & $0.2303$ & $0.7233$ & $0.0555$ & $0.4192$ & $1.2580$ & $1.2100$\\
          $\beta/H$ & $4968.80$ & $2185.68$ & $33929.11$ & $8215.37$ & $3673.07$ & $9404.39$\\
          $g^\ast$ & $10.0123$ & $10.6832$ & $6.9376$ & $10.5685$ & $10.6562$ & $10.6536$\\
          $U^{1/4}_0/({\rm MeV})$ & $0.3343$ & $0.8652$ & $0.0552$ & $0.4829$ & $2.1094$ & $0.8716$\\
          $v_w$ & $0.9734$ & $0.9488$ & $0.9117$ & $0.9424$ & $0.9844$ & $0.8209$\\
          \hline
        \end{tabular}
        }
        \caption{The benchmark points of PBH from FOPT with mass distribution.
        }
        \label{table:2}
    \end{table}
The plot of number density versus $M_{\rm PBH}$ of each {\bf BP} is shown in Fig.\ref{fig:5}.
Although the mass peaks fall within $10^{17}\,{\rm g}\sim10^{19}\,{\rm g}$,
the mass of PBH that the Hawking evaporation contribute the most is around $5\times 10^{14}\,{\rm g}$, in which the lifetime of PBH is comparable with age of the Universe,
result in the associate gamma-ray spectra peak at energy $\sim \mathcal{O}(100\,{\rm MeV})$, where EGERT/FermiLAT set stringent upper bounds.
The extragalactic gamma-ray spectrum of the mass distributions is shown in Fig.\ref{fig:6}.
Only {\bf BP-5a}  and {\bf BP-6a} are slightly over the FermiLAT and EGRET constraints.
Although {\bf BP-1a} to {\bf BP-4a} haven't been ruled out by the present experiments,
their number densities are too small to produce sufficient 511 keV flux.
The 511 keV gamma-ray flux of {\bf BP-6a} is shown in Fig.\ref{fig:7}.
Comparing Fig.~\ref{fig:7} with Fig.~\ref{fig:1}, we see that the contribution of {\bf BP-6a} to 511 keV fits quite well to INTEGRAL data points,
but its extragalactic gamma-ray flux exceeds the EGRET's upper limit in Fig.~\ref{fig:6}.
From Fig.~\ref{fig:5}, we know the number density of {\bf BP-6a} is higher than
other benchmark points by several order of magnitudes,
so that, except {\bf BP-6a}, no other points is able to produce sufficient 511 keV flux
comparing to the disk contribution.
As a result, after including the PBH mass function, we could not find proper benchmark point satisfying gamma-ray constraints and explaining 511 keV excess.
\bigskip

\section{Conclusion}
\label{sec:conclusion}

For the case of monochromatic distribution,
PBH with mass $1.0\times 10^{-17} \lesssim M_{\rm PBH}/M_\odot
\lesssim 8.0\times 10^{-17}$ and fractional abundance
$10^{-3} \lesssim f_{\rm PBH} \lesssim 1.0$ is favored to explain
the galaxy center 511 keV line excess.
%
The meteoritical mass PBH can be efficiently produced from the scenario that the cosmological FOPT aggregates dark fermions into FB as an intermediate state
in which the Fermi pressure and gravity reach to an equilibrium.
Subsequently, FB becomes unstable and collapses to form PBH due to the domination of attractive Yukawa interaction as the temperature cools down.
There exists several constraints from cosmological and astrophysical observations on this scenario,
and we selected the parameter points satisfying
the big bang nucleosynthesis $\Delta N_{\rm eff}\leq 0.5$
and upper bounds of gamma-ray flux from EGERT/COMPTEL/FermiLAT/INTEGRAL/SPI.
Setting the $\chi$ asymmetry $\eta_{\rm DM}\simeq \mathcal{O}(10^{-10})$,
the phase transition associating with vacuum energy
$\mathcal{O}(1)\lesssim B^{1/4}/{\rm MeV} \lesssim \mathcal{O}(100)$
produces the desired monochromatic PBH mass and abundance for 511 keV excess.
On the other hand, this scenario predicts correlated signals of extragalactic $\mathcal{O}({\rm MeV})$ gamma-ray and GW spectrum with peak frequency
from $10^{-5}$ Hz to $10^{-4}$ Hz.
In the future, these signatures will be either confirmed or ruled out by AMEGO/e-ASTROGAM/COSI/XGIX-THESUS
and $\mu$Ares.
Finally,  we try to include the PBH mass function by considering the radius distribution of false vacuum bubbles during FOPT. However, the mass function trends to populate at low PBH mass, which hampers the 511 keV excess from explanation.

\bigskip

\section*{Acknowledgment}
P.Y.Tseng is supported in part by the National Science and Technology Council with
Grant No. NSTC-111-2112-M-007-012-MY3.
Y.M.Yeh is supported in part by Ministry of Education with Grant No. 111J0382I4.

\bigskip


\end{document}